\documentclass[aps,prd,onecolumn,eqsecnum,amsmath,nofootinbib,preprintnumbers]{revtex4}%
\newcounter{example}[section]
\newenvironment{example}[1][]{\refstepcounter{example}\par\medskip
	\noindent \textbf{Example~\theexample. #1} \rmfamily}{\medskip}

\usepackage{amsmath}
\usepackage{color,graphicx,float,subfigure}
\usepackage{amsfonts,amssymb,theorem,mathrsfs,times}
\usepackage{bm}
\usepackage{amsmath}
{\theorembodyfont{\upshape}
	}
{\theorembodyfont{\upshape}
	}
{\theorembodyfont{\upshape}
	}
{\theorembodyfont{\upshape}
	\newtheorem{dn}{Definition}}
{\theorembodyfont{\upshape}
	}
{\theorembodyfont{\upshape}
	}

\newcommand{\dalm}{\kern1pt\vbox{\hrule height 0.9pt\hbox{\vrule width
			0.9pt\hskip 2.5pt\vbox{\vskip 5.5pt}\hskip 3pt\vrule width
			0.3pt}\hrule height 0.3pt}\kern1pt}

\begin{document}
	\title{Quasi-local studies of the particle surfaces and their stability in general spacetimes}
	
	%
	
	\author{ Yong Song\footnote{e-mail
			address: syong@cdut.edu.cn}}
	
	\author{Chuanyu Zhang\footnote{e-mail
			address: zhangchuanyu10@cdut.edu.cn (corresponding author)}}

	
	\affiliation{
	College of Mathematics and Physics\\
	Chengdu University of Technology, Chengdu, Sichuan 610059,
	China}
	

	\date{\today}
	
	\begin{abstract}
	In this paper, enlightened by the definition of the photon surface given by Claudel, Virbhadra and Ellis, we give a quasi-local definition of the particle surface. From this definition, one can study the evolution of the circular orbits in general spacetime. Especially, we pointed out that this definition can be used to get the spherical circular orbits in stationary spacetimes which cannot be got by the definition of Claudel, Virbhadra and Ellis. Further, we give a condition to exclude the particle surface in spacetime without gravity. Simultaneously, we give a quasi-local definition of the stability of the particle surface in general spacetime. From this definition, one can get the evolution equation of the innermost stable circular orbit (ISCO) in general spacetime. To verify the correctness of these definitions, we studied the circular orbits in some special cases and the results are all consistent with the previous results.
	\end{abstract}
	

	\maketitle


\section{Introduction}
\begin{figure}[htb]
	\centering
	\includegraphics[width=2.5in]{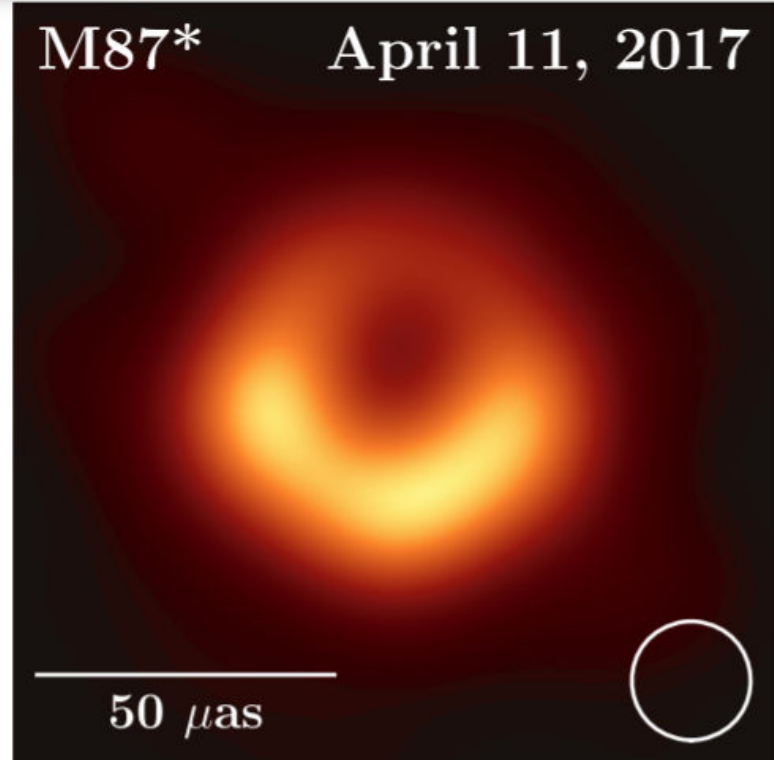}
	\caption{black hole image~\cite{EventHorizonTelescope:2019dse}:	The black hole image was taken in 2017 and published in 2019 after two years of data processing.}
	\label{bhphoto}
\end{figure}

Black holes are an important prediction of general relativity. Confirming the existence of black holes can further verify the correctness of general relativity. In 2019, the Event Horizon Telescope (EHT) Collaborations published the first image of the black hole at the center of the M87 galaxy~\cite{EventHorizonTelescope:2019dse,EventHorizonTelescope:2019ggy,EventHorizonTelescope:2019jan,EventHorizonTelescope:2019pgp,EventHorizonTelescope:2019ths,EventHorizonTelescope:2019uob}. As shown in FIG.\ref{bhphoto}.

In this image, one can clearly see a shadowed region surrounded by a halo. This shadow region is called black hole shadow, and the halo corresponds to the black hole accretion disk.  One can get a lot of information about the black hole by studying black hole shadow and accretion disk.  

Up to date, there are many studies on the black hole shadow~\cite{Grenzebach:2016,Cunha:2018acu,Virbhadra:2022ybp,Virbhadra:2022iiy} and the accretion disk~\cite{Abramowicz:2011xu}. The circular orbits of the spacetime are important for black hole shadow and accretion disk. For example, in Schwarzschild spacetime, on the one hand, the photon sphere, which is the location of the circular photon orbits, is the boundary of the black hole shadow~\cite{Synge:1966okc,Virbhadra:1999nm}. On the other hand, the timelike stable circular orbits determine the position of the accretion disk and the ISCO is the boundary of the accretion disk~\cite{Abramowicz:2011xu}. Here, we focus on the study of the circular orbits in the spacetime. There are mainly two ways to study the circular orbits. In the first method, one can solve the geodesic equations to get the circular orbits. This method has a wide range of applications. Such as, for null geodesics, this method can be used to study the photon sphere in static spacetimes~\cite{Hasse:2001by,Chakraborty:2011uj,Cardoso:2008bp}, the photon surface in dynamical spacetime~\cite{Mishra:2019trb,Koga:2022dsu} and the photon region in stationary spacetimes~\cite{Bardeen:1972fi,Teo:2003,Johannsen:2013vgc,Igata:2019pgb,Cederbaum:2019vwv}; for timelike geodesics, one can use this method to study the circular orbits in static spacetimes~\cite{Cornish:2003ig,Vieira:2013jga,Boonserm:2019nqq,Berry:2020ntz} and stationary spacetimes~\cite{Slany:2013ora,Stuchlik:2003dt,Teo:2020sey}. However, there are some practical problems in this traditional method which have been pointed out by~\cite{Cao:2019vlu}: (i). To grantee the separability of the geodesic equations, the systems which have been studied must have enough symmetries~\cite{Grenzebach:2016}. For real black holes, this condition is hard to satisfy. (ii). This method depends on the information at the infinity of the spacetime. This dependence means that we have to know the metric of the full spacetime, especially the future infinity of spacetime. This is impossible without additional assumptions. These problems inspire people to study the circular orbits in a different way. The second approach, which is called quasi-local study, has drawn some attention these years. For null geodesic, the first quasi-local definition of the photon surface is given by Claudel, Virbhadra and Ellis~\cite{Claudel:2000yi}. They defined a photon surface in a general spacetime as an umbilical timelike hypersurface. Based on this definition, they studied the photon surfaces in general spherically symmetric spacetimes. However, there are some obvious problems with this definition: (i). This definition is based on the null geodesic, so, it can not deal with the situation of timelike geodesic. (ii). The umbilical condition can not be used to study the photon region of stationary axisymmetric spacetimes. Up to date, there are some generalization studies of photon surface in stationary axisymmetric spacetimes. Such as, Yoshino et al. generalized the photon surface to be a loosely trapped surface~\cite{Shiromizu:2017ego} and (dynamically) transversely trapping surface~\cite{Yoshino:2017gqv,Yoshino:2019dty,Galtsov:2019bty}; Kobialko et al. generalized the photon surface to be a fundamental photon hypersurfaces and fundamental photon regions~\cite{Kobialko:2020vqf,Kobialko:2021aqg}. (iii).  This definition can not exclude the photon surface in spacetime without gravity. For example, it allows the existence of the photon surface in Minkowski spacetime. This problem has been solved by~\cite{Cao:2019vlu} which based on the geometry of a codimension-2 surface.

In this work, we want to give a generalized definition to over come the above three problems in the definition of the photon surface. Enlightened by the quasi-local definition of the photon sphere and photon surface~\cite{Claudel:2000yi,Cao:2019vlu}, we generalized the photon surface to the particle surface. Unlike the work studied the stability of the photon surface~\cite{Koga:2019uqd,Koga:2020akc}, we give a quasi-local definition of the stability of the particle surface in general spacetime. From this generalization, we studied the circular orbits in general spherical symmetric spacetimes and axisymmetric spacetimes. The results are all consistent with the previous results.

This paper is organized as follows: In section II, we will give
the definitions of a particle surface and the stability of the particle surface. In section III, based on these definitions, we will study the circular orbits in general static spherical symmetric spacetimes. In section IV, we will study the evolutions of the circular orbits in general spherical symmetric spacetimes. Especially, we will study the evolutions of the circular orbits in Vaidya spacetime. In section V, we will use the spherical orbits of Kerr spacetime and evolved spherical orbits of Kerr-Vaidya spacetime in the slow rotation limit as examples to show the validity of the definition 1. Section VI is devoted to the conclusion and discussion.

Convention of this paper: We choose the system of geometrized unit, i.e., set $G=c=1$. Also, we set the mass
of a massive point particle $m=1$ and use $M$ to denote the
mass of a black hole. We use the symbol $(\mathcal{M},\nabla_a,g_{ab})$ to denote a manifold $\mathcal{M}$ with metric $g_{ab}$ and covariant derivative operator $\nabla_a$, and ($g_{ab},\nabla_a$) satisfies the compatibility condition, i.e., $\nabla_a g_{bc}=0$. The abstract index formalism has been used to clarify some formulas or calculations. The curvature $R_{abcd}$ of the spacetime is defined by $R_{abcd}v^d=(\nabla_a\nabla_b-\nabla_b\nabla_a)v_c\,$ for an arbitrary tangent vector field $v^a$~\cite{Wald:1984rg}.


\section{The definition of a particle surface and the stability of the particle surface}\label{section2}
In this section, we will give the quasi-local definition of the particle surfaces and their stability in general spacetime. At first, let us give a brief review of the circular orbits in the Schwarzschild spacetime~\cite{Wald:1984rg}. The metric of the Schwarzschild spacetime in $\{t,r,\theta,\phi\}$ coordinates can be expressed as
\begin{eqnarray}
ds^2=-\bigg(1-\frac{2M}{r}\bigg)dt^2+\bigg(1-\frac{2M}{r}\bigg)^{-1}dr^2+r^2(d\theta^2+\sin^2\theta d\phi^2)
\end{eqnarray}
Due to the spherical symmetry of Schwarzschild spacetime, we can choose the equatorial plane, i.e., $\theta=\pi/2$. The circular photon orbits, which is usually called photon sphere or photon surface, is located at
\begin{eqnarray}
r=3M\;,
\end{eqnarray}
The null geodesics on the photon surface have the property that any null geodesic emitted in any tangent direction of the photon surface from any point on the photon surface will continue to propagate on the photon surface~\cite{Claudel:2000yi}. 

However, the case of timelike circular orbits is different. The locations of the timelike circular orbits are satisfies
\begin{eqnarray}
E=\frac{(r-2M)}{\sqrt{r(r-3M)}}\;,\quad L=\sqrt{\frac{Mr^2}{r-3M}}\;,
\end{eqnarray}
where $E$ is the conserved orbital energy and $L$ is the conserved orbital angular momentum of the timelike geodesic. And the tangent vector of the timelike circular orbits in the equatorial plane can be written as
\begin{eqnarray}
k^a=\bigg\{-\frac{E}{1-2M/r},0,0,\frac{L}{r^2}\bigg\}
\end{eqnarray}
The timelike circular orbits are located at the hypersurface that $r=\mathrm{constant}$, which will be called particle surface in this work. On each particle surface, the timelike geodesics with specified parameter have the property that it emitted in at least one tangent direction of the particle surface from any point on the particle surface will continue to propagate on the particle surface.

Enlightened by the above discussion and the quasi-local definition of the photon surface~\cite{Claudel:2000yi,Cao:2019vlu}, we give the following definitions:

\dn{\it{Let $(S,D_a,h_{ab})$ be a timelike hypersurface (or a subset of a timelike hypersurface) of $(\mathcal{M},\nabla_a,g_{ab})$. Let $k^a$ be the null or unit timelike tangent vector of $\gamma$ which is an affine null or timelike geodesic in $S$, i.e., it satisfies
\begin{eqnarray}
k^aD_ak^b=0\;.
\end{eqnarray} 
Let $v^a$ be a unit normal vector to $S$ and $K_{ab}$ be the second fundamental form on $S$. If for $\forall p\in S$, there exists at least one $\gamma\in S$ passing through $p$ and satisfies
\begin{eqnarray}
\label{Def1}
K_{ab}k^ak^b=0\;,
\end{eqnarray}
where $k^a$ can be timelike or null, and 
\begin{eqnarray}
\label{Def2}
2v^c\nabla_c(K_{ab}k^ak^b)\ne 0\;,
\end{eqnarray}
where $k^a$ is a null vector. Then, $S$ is called a (partial) particle surface.
}}

\dn{ \it {A (partial) particle surface is called stable if $2v^c\nabla_c(K_{ab}k^ak^b)\ge 0$, and unstabe if $2v^c\nabla_c(K_{ab}k^ak^b)<0$.}}

In definition 2, $k^a$ can be a timelike or a null vector and we have assumed that $v^a$ is out pointing. Roughly speaking, the out pointing requirement refers to a direction from the center of the system to infinity. The condition $2v^c\nabla_c(K_{ab}k^ak^b)=0$ for a timelike geodesic will determine the position of the innermost stable circular orbit of the spacetime.

\vspace{0.3cm}
To get a better understand of these definitions, we give some remarks:

(i). It should be noted that the spatial section of a particle surface may not be topologically closed, and it is called a partial particle surface in this work. For example, although the null particle surfaces exist in the Kerr spacetime at $r=\mathrm{constant}$, the angular range is limited as $-\theta_c\le\theta\le\theta_c$ from the work of Ref.~\cite{Teo:2003}.

(ii). Although definition 1 is quite general, it needs the spacetime have enough symmetries to solve eq.(\ref{Def1}). So, in this paper, we focus on the applications of the above definitions in spherical symmetric spacetimes and axisymmetric spacetimes. In spherical symmetric spacetimes and axisymmetric spacetimes, from the normalized condition
\begin{displaymath}
	k^ak_a = \left\{ \begin{array}{ll}
		-1\quad\quad & \textrm{timelike}\;,\\
		\\
		0\quad\quad & \textrm{null}\;,
	\end{array} \right.
\end{displaymath}
and the orthogonal condition
\begin{eqnarray}
k^av_a=0\;,\nonumber
\end{eqnarray}
and the condition of the (partial) particle surface
\begin{eqnarray}
K_{ab}k^ak^b=0\;,\nonumber
\end{eqnarray}
one can get the tangent vector $k^a$ and the equation of the (partial) particle surface. The detailed solution process will be developed in the following sections.

(iii). Based on the proof given by~\cite{Claudel:2000yi}, for any curve in $S$ with timelike or null tangent vector $k^a$, one has
\begin{eqnarray}
k^aD_ak^c=k^a\nabla_ak^c+(K_{ab}k^ak^b)v^c\;,\label{kdk}
\end{eqnarray}
where $D_a$ denotes covariant differentiation in $S$ with respect to $h_{ab}$. If the eq.(\ref{Def1}) in definition 1 is satisfied, the second term on the right of
eq.(\ref{kdk}) will be vanished, i.e., if $k^a$ is tangent to an affine timelike or null geodesic of $(S,h_{ab})$ then the term on the left of eq.(\ref{kdk}) also vanishes and so $k^a$ is tangent to an affine timelike or null geodesic of $(M,g_{ab})$.

(iv). The tangent vector $k^a$ satisfies the following condition:
\begin{displaymath}
	k^ak_a = \left\{ \begin{array}{ll}
		-1\quad\quad & \textrm{timelike}\;,\\
		\\
		0\quad\quad & \textrm{null}\;.
	\end{array} \right.
\end{displaymath}
If $k^ak_a=-1$, eq.(\ref{Def1}) can be used to get the stable or unstable circular orbits in general spacetime. If $k^ak_a=0$,  the definition 1 degenerates to the photon sphere in static spherical symmetric spacetime and the photon surface in dynamical spherical symmetric spacetime. The photon region in stationary axisymmetric spacetime can be formed by a collection of (partial) photon surface.

(v). In the general situation, from eq.(\ref{Def1}), one can not get the result that $\sigma_{ab}=0$, where $\sigma_{ab}$ is the trace-free part of $K_{ab}$. The umbilical condition, i.e., $\sigma_{ab}=0$, can be only used to get the photon surface in spherical symmetric spacetime~\cite{Claudel:2000yi}.

(vi). Eq.(\ref{Def2}) only need to be held for a null vector $k^a$, i.e., $k^ak_a=0$. As long as this condition is not satisfied, we will say that there is no particle surface in this spacetime. For a timelike geodesic, it can have continuous solutions of eq.(\ref{Def1}) and will have inhomogeneity along $v^a$. So, eq.(\ref{Def2}) will not vanish in general, even in the spacetime which is the absence of gravity, such as the Minkowski spacetime in the next section.

(vii). Eq.(\ref{Def2}) can be used to exclude the particle surface in the spacetime without gravity. It should be noted that our condition is based on a hypersurface, not a codimension-2 surface~\cite{Cao:2019vlu}. But it can not be used to exclude the asymptotically null photon surface in the spacetime with inhomogeneity of the gravitational field. Such as, the asymptotically null photon surface in Schwarzschild spacetime,
\begin{eqnarray}
\label{anps}
\bigg(\frac{dr}{dt}\bigg)^2=\bigg(1-\frac{2M}{r}\bigg)^2\bigg[1-\frac{b^2}{r^2}\bigg(1-\frac{2M}{r}\bigg)\bigg]\;,
\end{eqnarray}
where $b=L/E$ is the impact parameter. The solutions of eq.(\ref{anps}) also satisfy eq.(\ref{Def1}), and they can not be excluded by eq.(\ref{Def2}).

(viii). The factor ``2" in eq.(\ref{Def2}) or definition 2 is not important. From the following section we will know that $2v^c\nabla_{c}(K_{ab}k^ak^b)$ in definition 2 is exactly the second derivative of the effective potential in static spherical symmetric spacetime, and the factor $2$ is a proportionality coefficient. So, $2v^c\nabla_{c}(K_{ab}k^ak^b)$ can be used to characterize the stability of the (partial) particle surface.


\section{Static spherical symmetric spacetimes}\label{sections3}
In this section, we consider the circular orbits in general static spherical symmetric spacetimes. In the $\{t,r,\theta,\phi\}$ coordinates, the metric of the general static spherical symmetric spacetimes can be written as
\begin{eqnarray}
ds^2=-F(r)dt^2+H(r)dr^2+r^2(d\theta^2+\sin^2\theta d\phi^2)
\end{eqnarray}
where $F$ and $H$ are functions of $r$. Considering the untraped region, we have $F(r)>0$ and $H(r)>0$. In static spacetime, the circular orbits is not evolved, and they satisfy the condition that $r_o=\mathrm{constant}$, where $r_o$ is the location of the circular orbit. Then, we can get the normal vector $v^a$ as follows
\begin{eqnarray}
v^a=\frac{1}{\sqrt{H(r)}}\bigg(\frac{\partial}{\partial r}\bigg)^a\;,
\end{eqnarray}
Because the system has the spherical symmetry, we can choose the equatorial plane, i.e., $\theta=\pi/2$. And we can suppose the component of the tangent vector of the circular orbit to be $k^a=\{k^0,k^1,0,k^3\}$, where $k^0\;,k^1\;,k^3\;$ are only functions of $r$. From the orthogonal condition that $k^av_a=0$, we have
\begin{eqnarray}
k^1\sqrt{H(r_o)}=0\;,
\end{eqnarray}
Then we have $k^1=0$. So, we can choose $k=\{k^0,0,0,k^3\}$. From eq.(\ref{Def1}), we get
\begin{eqnarray}
	\label{kakbnablavb-1}
	2r_o^2(k^3)^2-r_oF'(r_o)(k^0)^2=0\;,
\end{eqnarray}
where $F'(r_o)=\partial F(r)/\partial r\mid_{r_o}$. To parameterize the circular orbit, we introduce the conserved orbital energy and the conserved orbital angular momentum which defined as
\begin{eqnarray}
	\label{E-1}
	&&E_o=k_a\bigg(\frac{\partial}{\partial t}\bigg)^a=-F(r_o)k^0\;,\\
	\label{L-1}
	&&L_o=k_a\bigg(\frac{\partial}{\partial \phi}\bigg)^a=r_o^2k^3\;.
\end{eqnarray}

\subsection{timelike geodesic}

For a timelike circular orbit, from the normalized condition of the timelike geodesic, i.e., $k^ak_a=-1$, we have
\begin{eqnarray}
\label{kaka-1}
-F(r_o)(k^0)^2+r_o^2(k^3)^2=-1\;,
\end{eqnarray}
Then, combining the eq.(\ref{kakbnablavb-1}), (\ref{E-1}), (\ref{L-1}) and (\ref{kaka-1}), we can get
\begin{eqnarray}
\label{tlEL1}
E_o^2=\frac{2F^2(r_o)}{2F(r_o)-r_oF'(r_o)}\;,\quad L_o^2=\frac{r_o^3F'(r_o)}{2F(r_o)-r_oF'(r_o)}\;,
\end{eqnarray}
which is consistent with the result in~\cite{Cardoso:2008bp}. From definition 2, we have
\begin{eqnarray}
\label{V2tl}
2v^c\nabla_c(K_{ab}k^ak^b)=\frac{2}{F(r)H(r)}\frac{-3F(r)F'(r)/r+2[F'(r)]^2-F(r)F''(r)}{2F(r)-rF'(r)}\;,
\end{eqnarray}
which is the second derivative of the effective potential of the timelike geodesic in~\cite{Cardoso:2008bp}\footnote{It needs to be noticed that $F(r)$ and $H(r)$ respectively correspond to the $f(r)$ and $1/g(r)$ in~\cite{Cardoso:2008bp} and there is minus sign bewteen the eq.(\ref{V2tl}) and the result in~\cite{Cardoso:2008bp} because of the difference of the signature of the metric. This situation also exist in eq.(\ref{V2nl}).}. It is well-known that ISCO satisfies that the second derivative of the effective potential is zero~\cite{Tsupko:2016bpn,Berenstein:2020vlp}. So, at the location of ISCO, eq.(\ref{Def2}) can be zero.

\subsection{null geodesic}
The circular orbit of a null geodesic in static spherical symmetric  spacetime is usually called a photon sphere. So, we will substitute $r_{ps}$ for $r_o$ to represent the location of a null circular orbit. For a photon sphere, from the normalized condition of the null geodesic, i.e., $k^ak_a=0$, we have
\begin{eqnarray}
	\label{kaka0}
	-F(r_{ps})(k^0)^2+r_{ps}^2(k^3)^2=0\;,	
\end{eqnarray}
Then, combining eq.(\ref{kakbnablavb-1}), (\ref{E-1}), (\ref{L-1}) and (\ref{kaka0}), we have
\begin{eqnarray}
\label{ss2.1tl}
	\frac{E_{ps}^2}{L_{ps}^2}=\frac{F(r_{ps})}{r_{ps}^2}\;,\quad 2F(r_{ps})=r_{ps}F'(r_{ps})\;,
\end{eqnarray}
which is consistent with~\cite{Cardoso:2008bp}. From eq.(\ref{Def2}), we have
\begin{eqnarray}
\label{V2nl}
	2v^c\nabla_c(K_{ab}k^ak^b)=\frac{L_{ps}^2}{r_{ps}^4F(r_{ps})H(r_{ps})}\big[2F(r_{ps})-r_{ps}^2F''(r_{ps})\big]\;,
\end{eqnarray}
which is the second derivative of the potential of the null geodesic in~\cite{Cardoso:2008bp}. 

\begin{example}
Here, we will use Minkowski spacetime as an example to illustrate how eq.(\ref{Def2}) excludes the particle surface in the spacetime which is the absence of gravity. The metric of the Minkowski spacetime in $\{t,r,\theta,\phi\}$ coordinate can be written as
\begin{eqnarray}
ds^2=-dt^2+dr^2+r^2(d\theta^2+\sin^2\theta d\phi^2)\;.
\end{eqnarray}
The unit normal vector $v^a$ of the particle surface is
\begin{eqnarray}
v^a=\frac{\dot{r}}{\sqrt{1-\dot{r}^2}}\bigg(\frac{\partial}{\partial t}\bigg)^a+\frac{1}{\sqrt{1-\dot{r}^2}}\bigg(\frac{\partial}{\partial r}\bigg)^a
\end{eqnarray}
where a $``\cdot"$ stands for the derivative with respect to coordinate time $``t"$. Considering the spherical symmetry of this spacetime, we study the circular orbits on the equatorial plane. Then, we suppose the component of the tangent vector of the circular orbit to be $k^a=\{k^0,k^1,0,k^3\}$, where $k^0\;,k^1\;,k^3\;$ are the functions of $\{t,r\}$. From the orthogonal condition, i.e., $k^av_a=0$, we have
\begin{eqnarray}
	\label{Minkv}
	\dot{r}_sk^0-k^1=0\;.
\end{eqnarray}
where $r_s$ represents $r_s(t)$ and is the location of the particle surface. From eq.(\ref{Def1}), we have
\begin{eqnarray}
	\label{Minkakakab}
k^ak^bK_{ab}=-\frac{\ddot{r}_s}{(1-\dot{r}_s^2)^{3/2}}(k^0)^2+\frac{\dot{r}\ddot{r}_s}{(1-\dot{r}_s^2)^{3/2}}k^0k^1+\frac{r(k^3)^2}{(1-\dot{r}^2)^{1/2}}=0\;.
\end{eqnarray}
\begin{itemize}
	\item timelike geodesic:
\end{itemize}
From the normalized condition of the timelike geodesic, i.e., $k^ak_a=-1$, we have
\begin{eqnarray}
	\label{Minkakatl}
	-(k^0)^2+(k^1)^2+r_s^2(k^3)^2=-1\;,
\end{eqnarray}
The orbital angular momentum of the timelike geodesic can be defined as
\begin{eqnarray}
\label{MinL}
L_s=r_s^2k^3\;.
\end{eqnarray}
So, combining eq.(\ref{Minkv}), (\ref{Minkakakab}), (\ref{Minkakatl}) and (\ref{MinL}), we get
\begin{eqnarray}
\label{Minrtl}
\ddot{r}_s=\frac{L_s^2(1-\dot{r}_s^2)}{r_s(L_s^2+r_s^2)}\;.
\end{eqnarray}
From definition 2, we have
\begin{equation}
\begin{aligned}
\label{Min2.2tl}
v^c\nabla_c(k^ak^bK_{ab})&=v^0\frac{\partial}{\partial t}\bigg(-\frac{r^3\ddot{r}+L^2(\dot{r}^2-1+r\ddot{r})}{r^3(1-\dot{r}^2)^{3/2}}\bigg)+v^1\frac{\partial}{\partial r}\bigg(-\frac{r^3\ddot{r}+L^2(\dot{r}^2-1+r\ddot{r})}{r^3(1-\dot{r}^2)^{3/2}}\bigg)\\
&=v^1\frac{\partial}{\partial r}\bigg(-\frac{r^3\ddot{r}+L^2(\dot{r}^2-1+r\ddot{r})}{r^3(1-\dot{r}^2)^{3/2}}\bigg)\\
&=\frac{3L^2(\dot{r}^2-1+r\ddot{r})}{r^4(1-\dot{r}^2)^{5/2}}\;.
\end{aligned}
\end{equation}
It should be pointed out that $r$ is a function of $``t"$ in eq.(\ref{Min2.2tl}). So, the second equal sign in eq.(\ref{Min2.2tl}) involves variational operation. Putting eq.(\ref{Minrtl}) into eq.(\ref{Min2.2tl}), we can easily get
\begin{eqnarray}
	v^c\nabla_c(k^ak^bk_{ab})=-\frac{3L_s^2}{r_s^2(L_s^2+r_s^2)(1+\dot{r}_s^2)^{3/2}}\ne 0\;.
\end{eqnarray}
So, eq.(\ref{Def2}) can be satisfied for timelike geodesic in Minkowski spacetime.


\begin{itemize}
	\item null geodesic:
\end{itemize}
From the normalized condition of the null geodesic, i.e., $k^ak_a=0$,, we have
\begin{eqnarray}
\label{Minkakan}
-(k^0)^2+(k^1)^2+r^2(k^3)^2=0\;,
\end{eqnarray}
Combining eq.(\ref{Minkv}), (\ref{Minkakakab}) and (\ref{Minkakan}), we get
\begin{eqnarray}
\label{Minddr}
\ddot{r}_s=\frac{1-\dot{r}_s^2}{r_s}\;.
\end{eqnarray}
Then, we can get the solution of the above equation is
\begin{eqnarray}
	\label{Minrn}
r_s^2=t^2+B^2\;,
\end{eqnarray}
where $B$ is an arbitrary constant factor. This is the timelike hyperboloid solution in~\cite{Claudel:2000yi}.
From eq.(\ref{Def2}), we have
\begin{equation}
\begin{aligned}
		\label{Min2.2n}
v^c\nabla_c(k^ak^bk_{ab})&=v^0\frac{\partial}{\partial t}\bigg(\frac{-r\ddot{r}+1-\dot{r}^2}{r\sqrt{1-\dot{r}^2}}(k^0)^2\bigg)+v^1\frac{\partial}{\partial r}\bigg(\frac{-r\ddot{r}+1-\dot{r}^2}{r\sqrt{1-\dot{r}^2}}(k^0)^2\bigg)\\
&=v^1(k^0)^2\frac{\partial}{\partial r}\bigg(\frac{-r\ddot{r}+1-\dot{r}^2}{r\sqrt{1-\dot{r}^2}}\bigg)\\
&=\frac{\dot{r}^2-1+r\ddot{r}}{r^2(1-\dot{r}^2)^{3/2}}v^1(k^0)^2\\
&=0\;.
\end{aligned}
\end{equation}
where we have used eq.(\ref{Minddr}) or (\ref{Minrn}).

From this example, we can clearly see how eq.(\ref{Def2}) excludes the particle surface in the spacetime which is the absence of gravity.
\end{example}

Conclusion of this section: The eq.(\ref{Def1}) of definition 1 can be used to get the circular orbits in static spherical spacetime. Definition 2 correspondings to the second derivative of the effective potential of the geodesic, so, it can be used to characterize the stability of the circular orbits in static spherical symmetric spacetimes. In more general situation, one may not define the effective potential of the system, but one can still use eq.(\ref{Def2}) to characterize the stability of the evolved circular orbits in general spacetimes.


\section{General spherical symmetric spacetime}
In this section, we will study the evolution of the circular orbits in general spherical symmetric spacetime. The metric of the general spherical symmetric spacetime in $\{t,r,\theta,\phi\}$ coordinate can be written as
\begin{eqnarray}
ds^2=-f(t,r)dt^2+g(t,r)dr^2+r^2(d\theta^2+\sin^2\theta d\phi^2)\;,
\end{eqnarray}
where $f$ and $g$ are functions of $\{t,r\}$. The unit normal vector $v^a$ of the particle surface can be written as~\cite{Claudel:2000yi}
\begin{eqnarray}
	v^a=\sqrt{\frac{g}{f}}\frac{\dot{r}}{\sqrt{f-g\dot{r}^2}}\bigg(\frac{\partial}{\partial t}\bigg)^a+\sqrt{\frac{f}{g}}\frac{1}{\sqrt{f-g\dot{r}^2}}\bigg(\frac{\partial}{\partial r}\bigg)^a\;.
\end{eqnarray}
where $f$ and $g$ represent $f(t,r_o(t))$ and $g(t,r_o(t))$, $r_o$ is a function of coordinate time $``t"$ and $``\cdot"$ stands for the derivative with respect to $``t"$. Considering the spherical symmetry of this spacetime, we study the circular orbits on the equatorial plane. Then, we suppose the component of the tangent vector of the circular orbit to be $k^a=\{k^0,k^1,0,k^3\}$, where $k^0\;,k^1\;,k^3\;$ are the functions of $\{t,r\}$. From the orthogonal condition, i.e., $k^av_a=0$, we have
\begin{eqnarray}
\label{dynamicalkv}
\dot{r}_ok^0-k^1=0\;.
\end{eqnarray}

\subsection{timelike geodesic}
For an evolved timelike circular orbit,from the normalized condition of the timelike geodesic, i.e., $k^ak_a=-1$,, we have
\begin{eqnarray}
\label{dynamicaltlkk}
-f(k^0)^2+g(k^1)^2+r_o^2(k^3)^2=-1\;,
\end{eqnarray}
To parametrize the timelike circular orbit  of the general dynamical spherical symmetric spacetime, we introduce the conserved orbital angular momentum of the timelike circular orbit which defined as
\begin{eqnarray}
	\label{dynamicalL}
	L_o=k_a\bigg(\frac{\partial}{\partial \phi}\bigg)^a=r_o^2k^3\;.
\end{eqnarray}
From eq.(\ref{Def1}) and considering eq.(\ref{dynamicalkv}), (\ref{dynamicaltlkk}) and (\ref{dynamicalL}), we have
\begin{equation}
\begin{aligned}
\label{dynamicaltl2.1}
k^ak^bK_{ab}&=\frac{(r_o^2+L_o^2)}{2r_o^3\sqrt{fg}(f-g\dot{r}_o^2)^{3/2}}\bigg\{2(f^2-\dot{r}_ofg)\frac{L_o^2}{r_o^2+L_o^2}-r_of^\prime(f-2\dot{r}_o^2g)-r_o\dot{r}_o[f(2\dot{g}+\dot{r}_og^\prime)-g(\dot{f}+\dot{g}\dot{r}_o^2)]-2r_ofg\ddot{r}_o\bigg\}\\
&=0\;,
\end{aligned}
\end{equation}
where $\dot{f}$ represents $\partial f(t,r)/\partial t\mid_{r_o(t),t}$ and $f^\prime$ represents $\partial f(t,r)/\partial r\mid_{r_o(t),t}$. Solving eq.(\ref{dynamicaltl2.1}), we have
\begin{eqnarray}
L_o^2=\frac{r_o^3[f(f^\prime+2\dot{g}\dot{r}_o+g^\prime\dot{r}_o^2+2g\ddot{r}_o)-g\dot{r}_o(\dot{f}+2f^\prime\dot{r}_o+\dot{g}\dot{r}_o^2)]}{2f(f-g\dot{r}_o^2)+gr_o\dot{r}_o(\dot{f}+2f^\prime \dot{r}_o+\dot{g}\dot{r}_o^2)-fr_o(f^\prime+2\dot{g}\dot{r}_o+g^\prime\dot{r}_o^2+2g\ddot{r}_o)}\;,
\end{eqnarray}
This is the evolution equation of the timelike circular orbits in general spherical symmetric spacetime and it is consistent with the result in~\cite{Song:2021ziq}.


\subsection{null geodesic}
The evolved circular photon orbit in dynamical spherical symmetric  spacetime is usually called a photon surface. So, we will substitute $r_{ph}$ for $r_o$ to represent the location of a null circular orbit. It should be noted that $r_{ph}$ represents $r_{ph}(t)$ and is a function of coordinate time $``t"$. For a photon surface, from the normalized condition of the null geodesic, i.e., $k^ak_a=0$, we have
\begin{eqnarray}
	\label{dynamicalnkk}
	-f(k^0)^2+g(k^1)^2+r_{ph}^2(k^3)^2=0\;,
\end{eqnarray}
From eq.(\ref{Def1}) and combining eq.(\ref{dynamicalkv}) and (\ref{dynamicalnkk}), we have
\begin{align}
\label{dynamicalkakbKabn}
k^ak^bK_{ab}&=\frac{L^2}{2r_{ph}^3(f-g\dot{r}_{ph}^2)^{3/2}}\sqrt{\frac{1}{fg}}\bigg\{2f^2+r_{ph}g\dot{r}_{ph}[\dot{f}+\dot{r}_{ph}(2f'+\dot{g}\dot{r}_{ph})]\nonumber\\
&-f[(f'+2\dot{g}\dot{r}_{ph}+g'\dot{r}_{ph}^2)r_{ph}+2g(\dot{r}_{ph}^2+\ddot{r}_{ph}r)]\bigg\}=0\;,
\end{align}
Solving eq.(\ref{dynamicalkakbKabn}), we get
\begin{eqnarray}
\ddot{r}_{ph}=\frac{2f^2+r_{ph}g\dot{r}_{ph}[\dot{f}+\dot{r}_{ph}(2f'+\dot{g}\dot{r}_{ph})]-f[2g\dot{r}^2_{ph}+r_{ph}(f'+2\dot{g}\dot{r}_{ph}+g'\dot{r}_{ph}^2)]}{2r_{ph}fg}\;.
\end{eqnarray}
This is the equation of the photon surface in general spherical symmetric spacetime.


\begin{example}
As an example, we study the evolution of the circular orbits in the in-going Vaidya spacetime. The metric of the 4-dimensional Vaidya spacetime in the in-going null coordinate $\{v,r,\theta,\phi\}$ can be written as~\cite{Vaidya:1951zza}
\begin{eqnarray}
ds^2=-\bigg(1-\frac{2M(v)}{r}\bigg)dv^2+2dvdr+r^2(d\theta^2+\sin^2\theta d\phi)\;,
\end{eqnarray}
where $M(v)$ is a freely specifiable function of $v$. The unit normal vector $v^a$ of the particle surface in Vaidya spacetime can be written as~\cite{Claudel:2000yi}
\begin{eqnarray}
v^a=\frac{1}{\sqrt{1-2M(v)/r-2\dot{r}}}\bigg(\frac{\partial}{\partial v}\bigg)^a+\frac{1-2M(v)/r-\dot{r}}{\sqrt{1-2M(v)/r-2\dot{r}}}\bigg(\frac{\partial}{\partial r}\bigg)^a\;.
\end{eqnarray}
where $``\cdot"$ stands for the derivative with respect to the coordinate time $``v"$. Because of the spherical symmetry of the system, we can study this question on the equatorial plane, i.e., $\theta=\pi/2$. Similarly, we can suppose the component of the tangent vector of the circular orbit to be $k^a=\{k^0,k^1,0,k^3\}$, where $k^0\;,k^1\;,k^3\;$ are functions of $\{v,r\}$. From the orthogonal condition that $k^av_a=0$, we have
\begin{eqnarray}
\label{vaidyakv}
\dot{r}_ok^0-k^1=0\;.
\end{eqnarray}
From eq.(\ref{Def1}), we have
\begin{eqnarray}
	\label{vaidya2.1}
	k^ak^bK_{ab}&=&\frac{-[\dot{M}(v)\dot{r}_o+(r_o-2M(v)-r_o\dot{r}_o)\ddot{r}_o](k^0)^2+(\dot{M}(v)+r_o\ddot{r}_o+\frac{M\dot{r}_o}{r_o})k^0k^1-\frac{M(v)}{r_o}(k^1)^2}{r_o(1-2M(v)/r_o-2\dot{r}_o)^{3/2}}\nonumber\\&+&\frac{2r_oM(v)k^0k^1-r_o^3[2M(v)-r_o+r_o\dot{r_o}](k^3)^2-[M(v)(r_o-2M(v)-r_o\dot{r}_o)+r_o^2\dot{M}(v)](k^0)^2}{r_o^3(1-2M(v)/r_o-2\dot{r}_o)^{1/2}}\nonumber\\
	&=&0\;.
\end{eqnarray}

\begin{itemize}
	\item timelike geodesic:
\end{itemize}
For a timelike geodesic, from the normalized condition that $k_ak^a=-1$, we have
\begin{eqnarray}
	\label{vaidyakk1}
	2k^0k^1-(k^0)^2\bigg(1-\frac{2M(v)}{r}\bigg)+r^2(k^3)^2=-1\;.
\end{eqnarray}
To parametrize the timelike circular orbit  of the Vaidya spacetime, we introduce the conserved orbital angular momentum of the timelike circular orbit which defined as
\begin{eqnarray}
	\label{vaidyaL}
	L_o=k_a\bigg(\frac{\partial}{\partial \phi}\bigg)^a=r_o^2k^3\;.
\end{eqnarray}
So, we have
\begin{eqnarray}
	\label{vaidyatlk3}
	k^3=\frac{L_o}{r_o^2}\;.
\end{eqnarray}
Combining eq.(\ref{vaidyakv}), (\ref{vaidyakk1}) and (\ref{vaidyatlk3}), we have
\begin{eqnarray}
	\label{vaidyatlk0}
	k^0=\sqrt{\frac{r_o^2+L_o^2}{r_o^2-2M(v)r_o-2r_o^2\dot{r}_o}}\;,
\end{eqnarray}
and
\begin{eqnarray}
	\label{vaidyatlk1}
	k^1=\sqrt{\frac{r_o^2+L_o^2}{r_o^2-2M(v)r_o-2r_o^2\dot{r}_o}}\dot{r}_o\;.
\end{eqnarray}
Here, we choose $k^a$ to be a furture-pointing vector. Putting eq.(\ref{vaidyatlk3}), (\ref{vaidyatlk0}) and (\ref{vaidyatlk1}) into eq.(\ref{vaidya2.1}), we have
\begin{eqnarray}
	L_o^2=\frac{2M^2(v)r_o^2+M(v)r_o^3(3\dot{r}_o-1)-r_o^4[\dot{M}(v)+r_o\ddot{r}_o]}{M(v)r_o(5-9\dot{r}_o)-6M^2(v)+r_o^2[r_o\ddot{r}_o+3\dot{r}_o-2\dot{r}_o^2+\dot{M}(v)-1]}\;.
\end{eqnarray}
which is consistent with the result in~\cite{Song:2021ziq}. To get the evolution equation of the ISCO in Vaidya spacetime, one needs to require $v^c\nabla_c(K_{ab}k^ak^b)=0$. Then, one can get
\begin{eqnarray}
\frac{\partial L_o^2}{\partial r_o}=0=\frac{\partial}{\partial r_o}\bigg(\frac{2M^2(v)r_o^2+M(v)r_o^3(3\dot{r}_o-1)-r_o^4[\dot{M}(v)+r_o\ddot{r}_o]}{M(v)r_o(5-9\dot{r}_o)-6M^2(v)+r_o^2[r_o\ddot{r}_o+3\dot{r}_o-2\dot{r}_o^2+\dot{M}(v)-1]}\bigg)\;,
\end{eqnarray}
which is consistent with the result in~\cite{Song:2021ziq}. For a more general situation, it is not hard to show definition 2 will give the same result as~\cite{Song:2021ziq}.

\begin{itemize}
	\item null geodesic:
\end{itemize}
For a photon surface in Vaidya spacetime, from the normalized condition of the null geodesic, i.e., $k_ak^a=0$, we have
\begin{eqnarray}
\label{vaidyakk0}
2k^0k^1-(k^0)^2\bigg(1-\frac{2M(v)}{r_{ph}}\bigg)+r_{ph}^2(k^3)^2=0\;.
\end{eqnarray}
From eq.(\ref{vaidyakv}) and (\ref{vaidyakk0}), we have
\begin{eqnarray}
\label{vaidyak1}
k^1=\dot{r}_{ph}k^0\;,
\end{eqnarray}
and
\begin{eqnarray}
\label{vaidyak3}
(k^3)^2=\frac{1-2M(v)/r_{ph}-2\dot{r}_{ph}}{r_{ph}^2}(k^0)^2\;.
\end{eqnarray}
Combining eq.(\ref{vaidyak1}), (\ref{vaidyak3}) and (\ref{vaidya2.1}), we have
\begin{eqnarray}
\ddot{r}_{ph}=\frac{1}{r_{ph}}\bigg[\bigg(1-\frac{3M(v)}{r_{ph}}\bigg)\bigg(1-\frac{2M(v)}{r_{ph}}-3\dot{r}_{ph}\bigg)-\dot{M}(v)+2\dot{r}_{ph}^2\bigg]\;.
\end{eqnarray}
This is the equation of the photon surface in Vaidya spacetime and it is consistent with the result in~\cite{Claudel:2000yi}.

\end{example}


\section{Kerr spacetime}
In this section, we will study the spherical orbits in Kerr spacetime to illustrate the validity of the definition in general stationary axisymmetric spacetimes. 

The metric of the $4$-dimensional Kerr spacetime in Boyer-Lindquist coordinates takes the following form
\begin{eqnarray}
	\label{kerr}
	ds^2=&-&(1-2Mr/\Sigma)dt^2-(4Mar\sin^2\theta/\Sigma)dtd\phi\nonumber\\
	&+&(\Sigma/\Delta)dr^2+\Sigma d\theta^2+(r^2+a^2+2Ma^2r\sin^2\theta/\Sigma)\sin^2\theta d\phi^2\;,
\end{eqnarray}
where $a$ is the angular momentum per unit mass of the black hole ($0\le a\le M$), and the functions $\Delta\;,\Sigma$ are defined as
\begin{eqnarray}
	\Delta&\equiv&r^2-2Mr+a^2\;,\\
	\Sigma&\equiv&r^2+a^2\cos^2\theta\;.
\end{eqnarray}
The spherical orbits are not evolved in general stationary axisymmetric spacetimes and satisfy $r=\mathrm{constant}$~\cite{Cederbaum:2019vwv}. So, the normal vector $v^a$ of the particle surface in Kerr spacetime can be expressed as
\begin{eqnarray}
	v^a=\sqrt{\frac{\Delta}{\Sigma}}\bigg(\frac{\partial}{\partial r}\bigg)^a\;,
\end{eqnarray}
The tangent vector of spherical orbits can be generally choosed as $k^a=\{k^0,k^1,k^2,k^3\}$. From the orthogonal condition, i.e. $k^av_a=0$, we have $k^1=0$. So, $k^a$ becomes $\{k^0,0,k^2,k^3\}$.

To obtain an explicit parameterization of the class of spherical orbits in the Kerr spacetime, we introduce the conserved orbital energy and orbital angular momentum as follows
\begin{eqnarray}
	\label{kerrE}
	&&E_o\equiv -k_a\bigg(\frac{\partial}{\partial t}\bigg)^a=\bigg(1-\frac{2Mr_o}{\Sigma_o}\bigg)k^0+\frac{2Mar_o\sin^2\theta}{\Sigma_o}k^3\;,\\
	\label{kerrL}
	&&L_o\equiv k_a\bigg(\frac{\partial}{\partial \phi}\bigg)^a=-\frac{2Mar_o\sin^2\theta}{\Sigma_o}k^0+(r_o^2+a^2+2Ma^2r_o\sin^2\theta/\Sigma_o)\sin^2\theta k^3\;.
\end{eqnarray}
where $\Sigma_o=r_o^2+a^2\cos^2\theta$. Simultaneously, we introduce the Carter's constant of the spherical orbit which can be defined as
\begin{eqnarray}
	\label{Q}
	Q_o=q_o-(L_o-aE_o)^2\;,
\end{eqnarray}
and $q_o$ is defined as
\begin{eqnarray}
	q_o=\kappa_{ab}k^ak^b\;,
\end{eqnarray}
where
\begin{eqnarray}
	\kappa_{ab}=\Sigma_o(l_an_b+n_al_b)+r_o^2g_{ab}\;,
\end{eqnarray}
and
\begin{eqnarray}
	&&l_a=-(dt)_a+\frac{\Sigma_o}{\Delta_o}(dr)_a+a\sin^2\theta(d\phi)_a\;,\\
	&&n_a=\frac{1}{2\Sigma_o}\bigg[-\Delta_o(dt)_a-\Sigma_o(dr)_a+a\sin\theta\Delta_o(d\phi)_a\bigg]\;.
\end{eqnarray}
where $\Delta_o=r_o^2-2Mr_o+a^2$. Then, we introduce the parameter 
\begin{eqnarray}
	\label{Phi}
	\Phi_o\equiv\frac{L_o}{E_o}=-\frac{2Mar_o\sin^2\theta-[(r_o^2+a^2)\Sigma_o+2Ma^2r_o\sin^2\theta]\chi\sin^2\theta }{\Sigma_o-2Mr_o+2Mar_o\chi\sin^2\theta}\;,
\end{eqnarray}
and
\begin{eqnarray}
	\label{QE2}
	\mathcal{Q}_o\equiv\frac{Q_o}{E_o^2}=\frac{\Sigma_o^2\Delta_o(1-a\chi \sin^2\theta)^2}{(\Sigma_o-2Mr_o+2Mr_oa\chi\sin^2\theta)^2}-(\Phi_o-a)^2\;.
\end{eqnarray}
where $\chi\equiv k^3/k^0$. Considering the eq.(\ref{Def1}), we have
\begin{equation}
\begin{aligned}
	\label{kerrkava}
k^ak^bK_{ab}&=-M(r_o^2-a^2\cos^2\theta)(k^0)^2+r_o\Sigma_o^2 (k^2)^2+2Ma(r_o^2-a^2\cos^2\theta)\sin^2\theta k^0k^3\\
&+\sin^2\theta\big[r_o\Sigma_o^2-Ma^2(r_o^2-a^2\cos^2\theta)\sin^2\theta\big](k^3)^2\\
&=0\;.
\end{aligned}
\end{equation}

\subsection{null geodesic}
The spatial section of the photon surface in Kerr spacetime is not topologically closed, and the collection of the photon surface forms the photon region in Kerr spacetime~\cite{Teo:2003}. For a spherical photon orbit, from the normalized condition of the null geodesic, i.e., $k^ak_a=0$, we have
\begin{eqnarray}
	\label{kerrkaka}
	-(\Sigma_o-2Mr_o)(k^0)^2+\Sigma_o^2 (k^2)^2-4Mr_oa\sin^2\theta k^0k^3+[(r_o^2+a^2)\Sigma_o+2Mr_o a^2\sin^2\theta]\sin\theta^2(k^3)^2=0\;.
\end{eqnarray}
Combining the eq.(\ref{Phi}), (\ref{QE2}), (\ref{kerrkava}) and (\ref{kerrkaka}), one can get
\begin{eqnarray}
&&\Phi_o=-\frac{r_o^3-3Mr_o^2+a^2r_o+Ma^2}{a(r_o-M)}\;,\\
&&\mathcal{Q}_o=-\frac{r_o^3(r_o^3-6Mr_o^2+9M^2r_o-4a^2M)}{a^2(r_o-M)^2}\;,
\end{eqnarray}
which are consistent with the results in~\cite{Teo:2003}.


\subsection{timelike geodesic}
For a spherical timelike orbit, from the normalized condition of the timelike geodesic, i.e., $k^ak_a=-1$, we have
\begin{eqnarray}
	\label{kerrkaka-1}
	\frac{(2Mr_o-\Sigma_o)(k^0)^2+\Sigma_o^2 (k^2)^2-4Mr_oa\sin^2\theta k^0k^3+[(r_o^2+a^2)\Sigma_o+2Mr_o a^2\sin^2\theta]\sin\theta^2(k^3)^2}{r_o^2+a^2\cos^2\theta}=-1\;.
\end{eqnarray}
Solving eq.(\ref{kerrE}) and (\ref{kerrL}), we get
\begin{eqnarray}
	\label{kerrk0}
&&k^0=E_o+\frac{2Mr_o[-aL_o+(r_o^2+a^2)E_o]}{\Delta_o \Sigma_o}\;,\\
\label{kerrk3}
&&k^3=\frac{2Mar_oE_o\sin^2\theta+(r_o^2-2M_or_o+a^2\cos^2\theta)L_o}{\Delta_o\Sigma_o\sin^2\theta}\;.
\end{eqnarray}
From eq.(\ref{Q}), we have
\begin{equation}
\begin{aligned}
	\label{tlQ}
Q=\Delta_o(k^0-ak^3\sin^2\theta)^2-(L_o-aE_o)^2\;.
\end{aligned}
\end{equation}
Putting eq.(\ref{kerrk0}) and (\ref{kerrk3}) into eq.(\ref{tlQ}), we get
\begin{eqnarray}
	\label{KerrQ}
Q_o=\frac{r_o[(2M-r_o)L_o^2-4MaE_oL_o+(r_o^3+2Ma^2+a^2r_o)E_o^2]}{\Delta_o}\;.
\end{eqnarray}
Combining eq.(\ref{kerrkava}), (\ref{kerrkaka-1}), (\ref{kerrk0}) and (\ref{kerrk3}), we have
\begin{eqnarray}
\label{kerrkakakab}
&&-4M^2r_o^3+(E_o^2-1)(r_o^5+2a^2r_o^3+a^4r_o)-a^2r_oL_o^2\nonumber\\
&&\qquad\qquad\qquad\qquad\qquad\qquad-M[(3E_o^2-4)r_o^4-a^2(L_o-aE_o)^2+2ar_o^2(aE_o^2-L_oE_o-2a)]=0\;,
\end{eqnarray}
Combining eq.(\ref{KerrQ}) and (\ref{kerrkakakab}), after some calculations, one can get the same result as~\cite{Teo:2020sey}
\begin{eqnarray}
&&E_o=\frac{r_o^3(r_o-2M)-a(aQ_o\mp\sqrt{Mr_o^5-Q_o(r_o-3M)r_o^3+a^2Q_o^2})}{r_o^2\sqrt{r_o^3(r_o-3m)-2a(aQ_o\pm\sqrt{Mr_o^5-Q_o(r_o-3M)r_o^3+a^2Q_o^2})}}\;,\\
&&L_o=-\frac{2Mar_o^3+(r_o^2+a^2)(aQ_o\mp\sqrt{Mr_o^5-Q_o(r_o-3M)r_o^3+a^2Q_o^2})}{r_o^2\sqrt{r_o^3(r_o-3m)-2a(aQ_o\pm\sqrt{Mr_o^5-Q_o(r_o-3M)r_o^3+a^2Q_o^2})}}\;.
\end{eqnarray}

\section{Kerr-Vaidya spacetime in the slow-rotation limit}
In this section, we will study the evolution of the spherical orbits of the Kerr-Vaidya spacetime in the slow-rotation limit.

For simplicity, we focus on the equatorial plane of the $4$-dimensional Kerr-Vaidya spacetime. In the slow rotation limit, the metric the Kerr-Vaidya spacetime on the equatorial plane can be expressed as~\cite{Mishra:2019trb,Murenbeeld:1970aq}
\begin{eqnarray}
	ds^2=-\bigg(1-\frac{2M(v)}{r}\bigg)dv^2+2dvdr-2adrd\phi-\frac{4M(v)a}{r}dvd\phi+r^2d\phi^2\;.
\end{eqnarray}
For an evolved circular orbits, one can suppose the component of the tangent vector of the evolved spherical orbits to be $k^a=\{k^0,k^1,k^2\}$, where $k^0\;,k^1\;,k^2\;$ are the functions of $\{v,r\}$. The unit normal vector $v^a$ of the particle surface can be written as
\begin{eqnarray}
	v^a=\sqrt{\frac{1-a\dot{\phi}}{1-2\dot{r}+a\dot{\phi}-2(1-a\dot{\phi})M(v)/r}}\bigg(\frac{\partial}{\partial v}\bigg)^a+\frac{1-\dot{r}-2(1-a\dot{\phi})M(v)/r}{\sqrt{[1-2\dot{r}+a\dot{\phi}-2(1-a\dot{\phi})M(v)/r](1-a\dot{\phi})}}\bigg(\frac{\partial}{\partial r}\bigg)^a\;,
\end{eqnarray}
where $``\cdot"$ stands for the derivative with respect to  the coordinate time $``v"$. From the orthogonal condition that $k^av_a=0$, we have
\begin{eqnarray}
\label{kerrvaidyakava}
(a\dot{\phi}_o-\dot{r}_o)k^0+(1-a\dot{\phi}_o)k^1+(\dot{r}_o-1)ak^2=0\;.
\end{eqnarray}
where $r_o$ and $\phi_o$ represent $r_o(v)$ and $\phi_o(v,r_o(v))$ respectively. From the eq.(\ref{Def1}) of definition 1, we have
\begin{equation}
\begin{aligned}
\label{kerrvaidya2.1}
k^ak^bK_{ab}&=\frac{1}{r^3[(1-a\dot{\phi}_o)(1-2\dot{r}_o+a\dot{\phi}_o+2(a\dot{\phi}_o-1)M(v)/r_o)]^{3/2}}\bigg\{r_o^3\bigg[-\frac{(k^0)^2}{r_o}[\dot{M}(v)(\dot{r}_o-a\dot{\phi}_o)(1-a\dot{\phi}_o)^2\\
&+(r_o\dot{r}_o-r_o+2M(v)-2aM(v)\dot{\phi}_o)(a\dot{\phi}_o\ddot{r}_o-\ddot{r}_o+a\ddot{\phi}_o-a\dot{r}_o\ddot{\phi}_o)]-\frac{k^0k^1}{r_o^2}\bigg(M(v)(1-a\dot{\phi}_o)^2(a\dot{\phi}_o-\dot{r}_o)\\
&+r_o\big[\dot{M}(v)(a\dot{\phi}_o-1)^3+r_o(1-a\dot{\phi}_o)^2\ddot{r}_o+ar_o(1-\dot{r}_o)(1-a\dot{\phi}_o)\ddot{\phi}_o\big]\bigg)+\frac{(a\dot{\phi}_o-1)^3M(v)(k^1)^2}{r_o^2}\\
&-\frac{k^0k^2}{r_o}\bigg(a\dot{M}(1-\dot{r}_o)(1-a\dot{\phi}_o)^2+a(ar_o\dot{\phi}_o-r_o\dot{r}_o-2M(v)+2aM(v)\dot{\phi}_o)(a\ddot{r}_o\dot{\phi}_o-\ddot{r}_o+a\ddot{\phi}_o-a\dot{r}_o\ddot{\phi}_o)\bigg)\\
&+\frac{aM(v)(1-\dot{r}_o)(1-a\dot{\phi}_o)^2k^1k^2}{r_o^2}\bigg]-\bigg[r_o^3(k^2)^2\big[2(1-a\dot{\phi}_o)M(v)-(1-\dot{r}_o)r_o\big]+2ar_oM(v)(1-a\dot{\phi}_o)k^1k^2\\
&+2aM(v)\big[2M(v)(1-a\dot{\phi}_o)-r_o(1-\dot{r}_o)\big]k^0k^2-2r_oM(v)(1-a\dot{\phi}_o)k^0k^1+\big[2(a\dot{\phi}_o-1)M(v)^2\\
&
+(1-\dot{r}_o)rM(v)+(1-a\dot{\phi}_o)r_o^2\dot{M}(v)\big](k^0)^2\bigg](1-a\dot{\phi}_o)\bigg(1-2\dot{r}_o+a\dot{\phi}_o+2(a\dot{\phi}_o-1)M(v)/r_o\bigg)\bigg\}\;.
\end{aligned}
\end{equation}
\subsection{null geodesic}
For an arbitrary null geodesic, we have $ds^2=0$. Then, we can get the following result
\begin{eqnarray}
r^2\dot{\phi}^2-\bigg(2a\dot{r}+\frac{4aM(v)}{r}\bigg)\dot{\phi}+2\dot{r}-\bigg(1-\frac{2M(v)}{r}\bigg)=0\;,
\end{eqnarray}
Solving the above equation, we get
\begin{eqnarray}
\dot{\phi}=\pm\frac{1}{r}\sqrt{1-\frac{2M(v)}{r}-2\dot{r}}+O(a)\;,
\end{eqnarray}
Here, we can only consider the lowest order of $a$. So, we have
\begin{eqnarray}
	\label{kerrvaidyaphi}
\dot{\phi}_o=\pm\frac{1}{r_o}\sqrt{1-\frac{2M(v)}{r_o}-2\dot{r}_o}+O(a)\;,
\end{eqnarray}
From the normalizing condition of $k^a$, i.e., $k^ak_a=0$, we have
\begin{eqnarray}
	\label{kerrvaidyakaka}
	(2M(v)-r_o)(k^0)^2+(2r_ok^1-4aM(v)k^2)k^0+r_o(r_o^2k^2-2ak^1)k^2=0\;.
\end{eqnarray}
Combining eq.(\ref{kerrvaidyakava}), (\ref{kerrvaidyakaka}), (\ref{kerrvaidya2.1}) and (\ref{kerrvaidyaphi}), we get
\begin{equation}
\begin{aligned}
\ddot{r}_o&=-\frac{3\dot{r}_o}{r_o}-\frac{\dot{M}(v)}{r_o}+\frac{9M(v)\dot{r}_o}{r_o^2}+\frac{2\dot{r}_o^2}{r_o}+\frac{1}{r_o}-\frac{5M(v)}{r_o^2}+\frac{6M(v)^2}{r_o^3}\\
&\mp\frac{6aM(v)}{r_o^3}\bigg[(2M(v)-r_o+2r_o\dot{r}_o)\sqrt{\frac{r_o-2M(v)-2r_o\dot{r}_o}{r_o^3}}\bigg]+O(a^2)\;.
\end{aligned}
\end{equation}
where ``$+$" correspond to the orbital angular momentum associated with ``direct" circular orbit and ``$-$" corresponds to the ``retrograde" circular orbit. This is the evolution equation of the circular photon orbits at the equatorial plane of the Kerr-Vaidya spacetime in the slow rotation limit, and it is consistent with the result in~\cite{Mishra:2019trb}.

\subsection{timelike geodesic}
For an arbitrary timelike geodesic, we have
\begin{eqnarray}
\label{kerrvaidyatl-1}
-\bigg(1-\frac{2M(v)}{r}\bigg)\bigg(\frac{dv}{d\tau}\bigg)^2+2\frac{dv}{d\tau}\frac{dr}{d\tau}-2a\frac{dr}{d\tau}\frac{d\phi}{d\tau}-\frac{4M(v)a}{r}\frac{dv}{d\tau}\frac{d\phi}{d\tau}+r^2\bigg(\frac{d\phi}{d\tau}\bigg)^2=-1
\end{eqnarray}
where $\tau$ is the affine parameter of the timelike geodesic, and the conserved orbital angular momentum can be defined as
\begin{eqnarray}
\label{kerrvaidyaL}
L=-\bigg(a\dot{r}+\frac{2aM(v)}{r}\bigg)\frac{dv}{d\tau}+r^2\frac{d\phi}{d\tau}\;,
\end{eqnarray}
Combining eq.(\ref{kerrvaidyatl-1}) and (\ref{kerrvaidyaL}), we get
\begin{eqnarray}
\dot{\phi}=\pm\frac{\sqrt{r-2M(v)-2r\dot{r}}}{r^{3/2}}\sqrt{\frac{L^2}{L^2+r^2}}+O(a)\;,
\end{eqnarray}
Here, we only consider the lowest order of $a$. So, we have
\begin{eqnarray}
	\label{kerrvaidyadphi-1}
	\dot{\phi}_o=\pm\frac{\sqrt{r_o-2M(v)-2r_o\dot{r}_o}}{r_o^{3/2}}\sqrt{\frac{L_o^2}{L_o^2+r_o^2}}+O(a)\;,
\end{eqnarray}
Considering an evolved timelike circular orbits of Kerr-Vaidya spacetime in the slow rotation limit, from the normalizing condition of $k^a$, i.e., $k^ak_a=-1$, we have
\begin{eqnarray}
	\label{kerrvaidyakaka-1}
	(2M(v)-r_o)(k^0)^2+(2r_ok^1-4aM(v)k^2)k^0+r_o(r_o^2k^2-2ak^1)k^2=-r_o\;.
\end{eqnarray}
and the corresponding angular momentum can be expressed as
\begin{eqnarray}
\label{kerrvaidyaL-1}
L_o=-\frac{2a M(v)}{r_o}k^0-ak^1+r_o^2k^2\;.
\end{eqnarray}
Combining eq.(
\ref{kerrvaidyakava}), (\ref{kerrvaidya2.1}), (\ref{kerrvaidyakaka-1}), (\ref{kerrvaidyadphi-1}) and (\ref{kerrvaidyaL-1}), we have
\begin{eqnarray}
	&&L_o^2=\frac{2M^2(v)r_{o}^2+M(v)r^3_{o}(3\dot{r}_{o}-1)-r^4_{o}[\dot{M}(v)+r_{o}\ddot{r}_{o}]}{M(v)r_{o}(5-9\dot{r}_{o})-6M^2(v)+r_{o}^2(r_{o}\ddot{r}_{o}+3\dot{r}_{o}-2\dot{r}^2_{o}+\dot{M}(v)-1)}\nonumber\\
	&&\pm \frac{6M(v)[2M(v)+r_o(2\dot{r}_o-1)]^2\sqrt{2M(v)+r_o(\dot{r}_o-1)}\sqrt{2M^2(v)r_{o}+M(v)r^2_{o}(3\dot{r}_{o}-1)-r^3_{o}[\dot{M}(v)+r_{o}\ddot{r}_{o}]}}{[M(v)r_{o}(5-9\dot{r}_{o})-6M^2(v)+r_{o}^2(r_{o}\ddot{r}_{o}+3\dot{r}_{o}-2\dot{r}^2_{o}+\dot{M}(v)-1)]^{2}}a\nonumber\\
	&&+O(a^2)\;,
\end{eqnarray}
This is the evolution equation of the timelike circular orbits at the equatorial plane of the Kerr-Vaidya spacetime in the slow rotation limit, and it is consistent with the result in~\cite{Song:2021ziq}.


\section{Discussion and conclusion}\label{conclusion}
In this paper, based on the definition of the photon surface given by~\cite{Claudel:2000yi,Cao:2019vlu}, we give a quasi-local definition of the (partial) particle surface in general spacetime in definition 1 and a simple classification in definition 2. The new definition has three advantages over~\cite{Claudel:2000yi}. First, it can be used to get the timelike circular orbits in general spacetime. Second, it can be used to study the photon region of the Kerr-like spacetime. Third, it not allows for the existence of the particle surface in spacetime without gravity. The third point is a little bit different from~\cite{Cao:2019vlu} which is based on a codimension-2 surface, and our definition is based on a hypersurface. 

In definition 2, we give a quasi-local definition of the stability of the circular orbits. In the static spherical symmetric spacetime, this condition consistent with the second derivative of the effective potential of the system. In more general situation, one may not define the effective potential of the system. But, one can still use definition 2 to classify the stability of the circular orbits. One can use the condition that $v^c\nabla_c(K_{ab}k^ak^b)=0$ to get the evolution equation of the ISCO in general spacetime, and the result is consistent with~\cite{Song:2021ziq}.

Using definition 1, we studied the (partial) particle surface in general spherical symmetric spacetime and axisymmetric spacetime. The results obtained by our definition are all consistent with the previous results. Although we only studied the situation in 4-dimensional spacetime, our definition actually can be used in higher spacetimes and it is easy to generalize the previous studies to higher spacetimes.

When solving the evolution equation of the (partial) particle surface, one need give appropriate boundary conditions. But from our definition, one can not get the boundary condition. The similar question in the study of photon surface has been solved by~\cite{Cao:2019vlu} which is based on a codimension-2 surface. One may solve the boundary question in our definition by using the method based on a codimension-2 surface.

\section*{Acknowledgement}

We would like to thank Li-Ming Cao for his useful discussions and kindly helps. This work is supported by the National Natural Science Foundation of China (Grant No. 41930112, 91755215 and 11804238), and our research is also supported by Sichuan Science and Technology Program (Grant No. 2022YFG0317). 


\end{document}